\documentclass[epsf,12pt,russian]{article}

\usepackage{amssymb}

\usepackage[dvips]{graphicx}

\begin{document}

\title
{Verification of new identity for the Green functions in $N=1$
supersymmetric non-Abelian Yang--Mills theory with the matter
fields.}

\author{A.B.Pimenov \thanks{E-mail:$pimenov.phys.msu@mail.ru$},
K.V.Stepanyantz\thanks{E-mail:$stepan@phys.msu.ru$}}

\maketitle

\begin{center}
{\em Moscow State University, physical faculty,\\
department of theoretical physics.\\
$119992$, Moscow, Russia}
\end{center}

\begin{abstract}
We investigate a new identity for Green functions using the higher
covariant derivative regularization. It relates some coefficients
in the vertex function of the matter superfield, in which one of
external matter lines is not chiral. The calculation in the first
nontrivial order (for the two-loop vertex function) reveals that
the new identity is also valid for the non-Abelian Yang--Mills
theory with matter fields. The new identity is shown to appear
because three-loop integrals, defining the Gell-Mann--Low function
are factorized into integrals of total derivatives.
\end{abstract}


\section{Introduction.}
\hspace{\parindent}

Investigation of quantum corrections in supersymmetric theories is
an interesting and sometimes nontrivial problem. For example, in
theories with the $N=1$ supersymmetry it is possible to suggest
\cite{NSVZ_Instanton} a form of the $\beta$-function exactly to
all orders. One way of obtaining the exact $\beta$-function,
proposed in Ref. \cite{SD,SDYM}, is substituting the solution of
Slavnov--Taylor identities into the Schwinger--Dyson equations.
Then the exact $\beta$-function is obtained if we propose
existence of a new identity, relating the Green functions
\cite{SD,SDYM}. Due to this identity some contributions to the
Gell-Mann-Low function disappear starting from the three-loop
approximation.

The existence of the new identity is related with the interesting
observation \cite{ThreeLoop,2LoopYM}, which was made using the
higher derivative regularization \cite{Slavnov,Slavnov_Book} in
supersymmetric theories. All contributions to the Gell-Mann--Low
function appear to be integrals of total derivatives. Partially
this can be explained substituting solutions of Ward identities
into the Schwinger--Dyson equations. In the Abelian case a
straightforward summation of diagrams is also possible
\cite{Identity}. Nevertheless, there are new types of diagrams in
the non-Abelian case and a method, used in Ref. \cite{Identity},
is not already working. Therefore, in the non-Abelian case it is
necessary to verify the new identity agin. Such a verification is
made in this paper.

This paper is organized as follows.

In Sec. \ref{Section_SUSY} we recall basic information about the
$N=1$ supersymmetric Yang-Mills theory, the background field
method, and the higher derivatives regularization. A verification
of the new identity is made in Sec. \ref{Section_F_Calculation}. A
brief discussion of the results is given in the conclusion. Some
technical details are presented in the Appendix.


\section{$N=1$ supersymmetric Yang-Mills theory, background field method
and higher derivative regularization} \label{Section_SUSY}
\hspace{\parindent}

We will consider the $N=1$ supersymmetric Yang-Mills theory with
massless matter superfields, which in the superspace is described
by the action

\begin{eqnarray}\label{SYM_Action}
&& S = \frac{1}{4 e^2} \mbox{Re}\,\mbox{tr}\int
d^4x\,d^2\theta\,W_a C^{ab} W_b + \frac{1}{4}\int d^4x\,
d^4\theta\, \Big(\phi^+ e^{2V}\phi +\tilde\phi^+
e^{-2V^{t}}\tilde\phi\Big).
\end{eqnarray}

\noindent Here $\phi$ and $\tilde\phi$ are chiral matter
superfields, and $V$ is a real scalar superfield, which contains
the gauge field $A_\mu$ as a component. The superfield $W_a$ is a
supersymmetric analogue of the gauge field stress tensor. It is
defined by

\begin{equation}
W_a = \frac{1}{32} \bar D (1-\gamma_5) D\Big[e^{-2V}
(1+\gamma_5)D_a e^{2V}\Big].
\end{equation}

\noindent In our notation, the gauge superfield $V$ is expanded
over the generators of the gauge group $T^a$ as $V = e\, V^a T^a$,
where $e$ is a coupling constant. Generators of the fundamental
representation we normalize by the condition

\begin{equation}
\mbox{tr}(t^a t^b) = \frac{1}{2} \delta^{ab}.
\end{equation}

\noindent Action (\ref{SYM_Action}) is invariant under the gauge
transformations

\begin{equation}
\phi \to e^{i\Lambda}\phi;\qquad \tilde\phi \to
e^{-i\Lambda^t}\tilde\phi;\qquad e^{2V} \to e^{i\Lambda^+} e^{2V}
e^{-i\Lambda},
\end{equation}

\noindent where $\Lambda$ is an arbitrary chiral superfield.

For quantization of this model it is convenient to use the
background field method. The matter is that the background field
method allows constructing the effective action, which is
invariant under some background gauge transformations. In the
supersymmetric case it can formulated as follows \cite{West,
Superspace}: Let us make the substitution

\begin{equation}\label{Substitution}
e^{2V} \to e^{2V'} \equiv e^{\mbox{\boldmath${\scriptstyle
\Omega}$}^+} e^{2V} e^{\mbox{\boldmath${\scriptstyle \Omega}$}}
\end{equation}

\noindent in action (\ref{SYM_Action}), where
$\mbox{\boldmath${\Omega}$}$ is a background scalar superfield.
The obtained theory will be invariant under the background gauge
transformations

\begin{equation}\label{Background_Transformations}
V \to e^{iK} V e^{-iK}; \qquad e^{\mbox{\boldmath${\scriptstyle
\Omega}$}} \to e^{iK} e^{\mbox{\boldmath${\scriptstyle \Omega}$}}
e^{-i\Lambda};\qquad e^{\mbox{\boldmath${\scriptstyle \Omega}$}^+}
\to e^{i\Lambda^+} e^{\mbox{\boldmath${\scriptstyle \Omega}$}^+}
e^{-iK},
\end{equation}

\noindent where $K$ is a real superfield and $\Lambda$ is a chiral
superfield.

Let us construct the chiral covariant derivatives

\begin{equation}
\mbox{\boldmath$D$} \equiv e^{-\mbox{\boldmath${\scriptstyle
\Omega}$}^+} \frac{1}{2} (1+\gamma_5)D
e^{\mbox{\boldmath${\scriptstyle \Omega}$}^+};\qquad
\bar{\mbox{\boldmath$D$}} \equiv e^{\mbox{\boldmath${\scriptstyle
\Omega}$}} \frac{1}{2} (1-\gamma_5) D
e^{-\mbox{\boldmath${\scriptstyle \Omega}$}}.
\end{equation}

\noindent Acting on some field $X$, which is transformed as $X \to
e^{iK} X$, these covariant derivatives are transformed in the same
way. It is also possible to define the background covariant
derivative with a Lorentz index

\begin{equation}
\mbox{\boldmath$D$}_\mu \equiv - \frac{i}{4} (C\gamma^\mu)^{ab}
\Big\{\mbox{\boldmath$D$}_a,\bar{\mbox{\boldmath$D$}}_b\Big\},
\end{equation}

\noindent which will have the same property. It is easy to see
that after substitution (\ref{Substitution}) action
(\ref{SYM_Action}) will be

\begin{eqnarray}\label{Background_Action}
&& S = \frac{1}{2e^2}\mbox{tr}\,\mbox{Re}\,\int d^4x\,d^2\theta\,
\mbox{\boldmath$W$}^a \mbox{\boldmath$W$}_a - \frac{1}{64
e^2}\mbox{tr}\,\mbox{Re}\,\int d^4x\,d^4\theta\,\Bigg[16
\Big(e^{-2V}\mbox{\boldmath$D$}^a e^{2V}\Big)
\mbox{\boldmath$W$}_a
+\nonumber\\
&& + \Big(e^{-2V}\mbox{\boldmath$D$}^a e^{2V}\Big)
\bar{\mbox{\boldmath$D$}}^2 \Big(e^{-2V}\mbox{\boldmath$D$}_a
e^{2V}\Big) \Bigg],
\end{eqnarray}

\noindent where

\begin{equation}
\mbox{\boldmath$W$}_a = \frac{1}{32}
e^{\mbox{\boldmath${\scriptstyle \Omega}$}} \bar D (1-\gamma_5) D
\Big(e^{-\mbox{\boldmath${\scriptstyle \Omega}$}}
e^{-\mbox{\boldmath${\scriptstyle \Omega}$}^+} (1+\gamma_5) D_a
e^{\mbox{\boldmath${\scriptstyle \Omega}$}^+}
e^{\mbox{\boldmath${\scriptstyle \Omega}$}} \Big)
e^{-\mbox{\boldmath${\scriptstyle \Omega}$}},
\end{equation}

\noindent and the notation

\begin{eqnarray}\label{Derivatives_Notations}
&& \mbox{\boldmath$D$}^2 \equiv \frac{1}{2} \bar
{\mbox{\boldmath$D$}} (1+\gamma_5)\mbox{\boldmath$D$};\qquad\ \bar
{\mbox{\boldmath$D$}}^2 \equiv \frac{1}{2} \bar
{\mbox{\boldmath$D$}} (1-\gamma_5) \mbox{\boldmath$D$};\nonumber\\
&& \mbox{\boldmath$D$}^a \equiv \Big[\frac{1}{2}\bar
{\mbox{\boldmath$D$}} (1+\gamma_5)\Big]^a;\qquad
\mbox{\boldmath$D$}_a \equiv \Big[\frac{1}{2}(1+\gamma_5)
\mbox{\boldmath$D$}\Big]_a;\nonumber\\
&& \bar {\mbox{\boldmath$D$}}^a \equiv \Big[\frac{1}{2}\bar
{\mbox{\boldmath$D$}} (1 - \gamma_5)\Big]^a;\qquad \bar
{\mbox{\boldmath$D$}}_a \equiv \Big[\frac{1}{2}(1-\gamma_5)
\mbox{\boldmath$D$}\Big]_a
\end{eqnarray}

\noindent is used. Action of the covariant derivatives on the
field $V$ in the adjoint representation is defined by the standard
way.

We note that action (\ref{Background_Action}) is also invariant
under the quantum transformations

\begin{equation}\label{Quantum_Transformations}
e^{2V} \to e^{-\lambda^+} e^{2V} e^{-\lambda};\qquad
\mbox{\boldmath${\Omega}$} \to \mbox{\boldmath${\Omega}$};\qquad
\mbox{\boldmath${\Omega}$}^+ \to \mbox{\boldmath${\Omega}$}^+
\end{equation}

\noindent where $\lambda$ is a background chiral superfield, which
satisfies the condition

\begin{equation}
\bar{\mbox{\boldmath$D$}}\lambda = 0.
\end{equation}

\noindent Such a superfield can be presented in the form $\lambda
= e^{\mbox{\boldmath${\scriptstyle \Omega}$}} \Lambda
e^{-\mbox{\boldmath${\scriptstyle \Omega}$}} $, where $\Lambda$ is
a usual chiral superfield.

It is convenient to choose a regularization and gauge fixing so
that invariance (\ref{Background_Transformations}) will be
unbroken. We fix the gauge by adding

\begin{equation}\label{Gauge_Fixing}
S_{gf} = - \frac{1}{32 e^2}\,\mbox{tr}\,\int d^4x\,d^4\theta\,
\Bigg(V \mbox{\boldmath$D$}^2 \bar{\mbox{\boldmath$D$}}^2  V + V
\bar {\mbox{\boldmath$D$}}^2 \mbox{\boldmath$D$}^2 V\Bigg)
\end{equation}

\noindent to the action. In this case terms quadratic in the
superfield $V$ will have the simplest form:

\begin{equation}
\frac{1}{2 e^2}\mbox{tr}\,\mbox{Re}\int d^4x\,d^4\theta\, V
\mbox{\boldmath$D$}_\mu^2 V.
\end{equation}

\noindent Also \cite{West} it is necessary to add an action for
the Faddeev--Popov ghosts $S_{c}$ and an action for the
Nielsen--Kallosh ghosts. Because in this paper we will calculate a
contribution of the matter superfields, we are not interested in
the concrete form of these terms. The gauge fixing breaks the
invariance of the action under quantum gauge transformations
(\ref{Quantum_Transformations}), but there is a remaining
invariance under the BRST-transformations. The BRST-invariance
leads to the Slavnov--Taylor identities, which relate vertex
functions of the quantum gauge field and ghosts.

However, in order to simplify the calculations it is convenient to
choose a regularization so that it breaks the invariance under the
BRST-transformations. We will add the following term with the
higher covariant derivatives

\begin{eqnarray}\label{Regularized_Action}
&& S_{\Lambda} = \frac{1}{2 e^2}\mbox{tr}\,\mbox{Re}\int
d^4x\,d^4\theta\, V\frac{(\mbox{\boldmath$D$}_\mu^2)^{n+1}}{
\Lambda^{2n}} V
\end{eqnarray}

\noindent to the action. (A method, used here, is a slightly
different from the one, proposed in Ref. \cite{West_Paper}.)
Because the regularization is not invariant under the
BRST-transformations, it is necessary to use a special
renormalization scheme, which ensures that the Slavnov--Taylor
identities are satisfied in each order of the perturbation theory
due to some additional subtractions. Such a renormalization scheme
was proposed in Refs. \cite{Slavnov1,Slavnov2}, and generalized to
the supersymmetric case in Refs. \cite{Slavnov3,Slavnov4}.

It is important to note that the Gell-Mann--Low function is scheme
independent and does not depend on a regularization and a
renormalization prescription.

The generating functional is written as

\begin{eqnarray}\label{Generating_Functional}
&& Z[J,\mbox{\boldmath$\Omega$}] = \int D\mu\,\exp\Big\{i S + i
S_\Lambda + i S_{gf} + i S_{gh} + iS_{\phi_0}
+\nonumber\\
&& \qquad\qquad\qquad\qquad\qquad + i \int d^4x\,d^4\theta\,\Big(J
+ J[\mbox{\boldmath$\Omega$}] \Big)
\Big(V'[V,\mbox{\boldmath$\Omega$}] - {\bf V} \Big) \Big\},\qquad
\end{eqnarray}

\noindent where the superfield ${\bf V}$ is defined by

\begin{equation}\label{Background Field}
e^{2{\bf V}} \equiv e^{\mbox{\boldmath${\scriptstyle \Omega}$}^+}
e^{\mbox{\boldmath${\scriptstyle \Omega}$}},
\end{equation}

\noindent and $J[\mbox{\boldmath$\Omega$}]$ is an arbitrary
functional. $S_{gf}$ is gauge fixing action (\ref{Gauge_Fixing})
and $S_{gh} = S_c + S_B$ is the corresponding action for the
Faddeev--Popov and Nielsen--Kallosh ghosts. Moreover, we added
terms with additional sources $\phi_0$, which are written as

\begin{eqnarray}\label{New_Sources}
&& S_{\phi_0} = \frac{1}{4}\int d^4x\,d^4\theta\,\Big(\phi_0^+
e^{\mbox{\boldmath${\scriptstyle \Omega}$}^+} e^{2V}
e^{\mbox{\boldmath${\scriptstyle \Omega}$}} \phi + \phi^+
e^{\mbox{\boldmath${\scriptstyle \Omega}$}^+} e^{2V}
e^{\mbox{\boldmath${\scriptstyle \Omega}$}}
\phi_0 +\nonumber\\
&& \qquad\qquad\qquad\qquad + \tilde\phi_0^+
e^{-\mbox{\boldmath${\scriptstyle \Omega}$}^*} e^{-2V^t}
e^{-\mbox{\boldmath${\scriptstyle \Omega}$}^t} \tilde\phi +
\tilde\phi^+ e^{-\mbox{\boldmath${\scriptstyle \Omega}$}^*}
e^{-2V^t} e^{-\mbox{\boldmath${\scriptstyle \Omega}$}^t}
\tilde\phi_0 \Big).\qquad
\end{eqnarray}

\noindent Unlike the fields $\phi$ and $\tilde\phi$, the fields
$\phi_0$ and $\tilde\phi_0$ are not chiral. In principle, adding
such terms is not quite necessary, but it is convenient to
formulate the new identity in terms of variational derivatives
with respect to these sources.

Using the functional $Z[J,\mbox{\boldmath$\Omega$}]$ it is
possible to construct the generating functional for connected
Green functions

\begin{equation}
W[J,\mbox{\boldmath$\Omega$}] = -i\ln
Z[J,\mbox{\boldmath$\Omega$}] = - \int d^4x\,d^4\theta \Big(J +
J[\mbox{\boldmath$\Omega$}]\Big) {\bf V} + W_0\Big[J +
J[\mbox{\boldmath$\Omega$}],\mbox{\boldmath$\Omega$}\Big]\qquad
\end{equation}

\noindent and the corresponding effective action

\begin{equation}\label{Effective_Action_Definition}
\Gamma[V,\mbox{\boldmath$\Omega$}] = -\int d^4x\,d^4\theta\,\Big(J
{\bf V} + J[\mbox{\boldmath$\Omega$}] {\bf V}\Big) + W_0\Big[J +
J[\mbox{\boldmath$\Omega$}],\mbox{\boldmath$\Omega$}\Big] - \int
d^4x\,d^4\theta\,J V,
\end{equation}

\noindent where the sources should be expressed in terms of fields
using the equation

\begin{eqnarray}
&& V = \frac{\delta}{\delta J} W[J,\mbox{\boldmath$\Omega$}] = -
{\bf V} + \frac{\delta}{\delta J} W_0\Big[J +
J[\mbox{\boldmath$\Omega$}],\mbox{\boldmath$\Omega$}\Big].
\end{eqnarray}

In order to understand how $\Gamma[V,\mbox{\boldmath$\Omega$}]$ is
related with the ordinary effective action, we perform the
substitution $V \to V'$ in the generating functional $Z$. Then we
obtain

\begin{equation}
Z[J,\mbox{\boldmath$\Omega$}] = \exp\Big\{ -i \int
d^4x\,d^4\theta\,\Big(J + J[\mbox{\boldmath$\Omega$}]\Big) {\bf V}
\Big\} Z_0\Big[J + J[\mbox{\boldmath$\Omega$}],
\mbox{\boldmath$\Omega$}\Big],
\end{equation}

\noindent where

\begin{equation}
Z_0[J,\mbox{\boldmath$\Omega$}] = \int D\mu\,\exp\Big\{i S + i
S_\Lambda + i S_{gf} + i S_{gh} + i \int d^4x\,d^4\theta\,J V
\Big\}.
\end{equation}

\noindent Therefore,

\begin{eqnarray}
&& \Gamma[V,\mbox{\boldmath$\Omega$}] = W_0\Big[J +
J[\mbox{\boldmath$\Omega$}],\mbox{\boldmath$\Omega$}\Big] - \int
d^4x\,d^4\theta\,\Big(J[\mbox{\boldmath$\Omega$}] {\bf V} + J
\frac{\delta}{\delta J} W_0\Big[J +
J[\mbox{\boldmath$\Omega$}],\mbox{\boldmath$\Omega$}\Big]\Big).
\end{eqnarray}

\noindent Let us now set $V = 0$, so that

\begin{equation}\label{Phi_To_Zero}
{\bf V} = \frac{\delta}{\delta J} W_0\Big[J +
J[\mbox{\boldmath$\Omega$}],\mbox{\boldmath$\Omega$}\Big].
\end{equation}

\noindent and take into account that in this case the superfield
$K$ is nontrivially present only in gauge transformation
(\ref{Background_Transformations}) for the fields
$\mbox{\boldmath$\Omega$}$ and $\mbox{\boldmath$\Omega$}^+$, and
the only invariant combination is expression (\ref{Background
Field}). (It is invariant in a sense, that the corresponding
transformation law does not contain the superfield $K$.)
Therefore, if $V=0$, then we can set

\begin{equation}\label{K_Fixing}
\mbox{\boldmath$\Omega$} = \mbox{\boldmath$\Omega$}^+ = {\bf V}.
\end{equation}

\noindent In this case the effective action is

\begin{eqnarray}\label{Background_Gamma}
&& \Gamma[0,{\bf V}] = W_0\Big[J + J[{\bf V}],{\bf V}\Big] - \int
d^4x\,d^4\theta\,\Big(J + J[{\bf V}]\Big) \frac{\delta}{\delta J}
W_0\Big[J + J[{\bf V}], {\bf V}\Big].
\end{eqnarray}

\noindent and does not depend on the form of the functional
$J[\mbox{\boldmath$\Omega$}]$.

If the gauge fixing terms, ghosts, and the terms with higher
derivatives depended only on $V'$, expression
(\ref{Background_Gamma}) would coincide with the ordinary
effective action. However, the dependence on $V$,
$\mbox{\boldmath$\Omega$}$, and $\mbox{\boldmath$\Omega$}^+$ is
not factorized into the dependence on $V'$ in the proposed method
of renormalization and gauge fixing. According to Ref.
\cite{Kluberg1,Kluberg2} the invariant charge (and, therefore, the
Gell-Mann-Low function) is gauge independent, and the dependence
of the effective action on gauge can be eliminated by
renormalization of the wave functions of the gauge field, ghosts,
and matter fields. Therefore, for calculating the Gell-Mann--Low
function we may use the background gauge described above.

Nevertheless, generating functional (\ref{Generating_Functional})
is not yet completely constructed. The matter is that adding the
term with higher derivatives does not remove divergences from
one-loop diagrams. To regularize them, it is necessary to insert
the Pauli-Villars determinants in the generating functional
\cite{Slavnov_Book}:

\begin{equation}\label{PV_Insersion}
\prod\limits_i \Big(\det PV(V,{\bf V},M_i)\Big)^{c_i},
\end{equation}

\noindent where the coefficients $c_i$ satisfy conditions

\begin{equation}
\sum\limits_i c_i = 1;\qquad \sum\limits_i c_i M_i^2 = 0.
\end{equation}

\noindent The Pauli--Villars fields are constructed for the
quantum gauge field, ghosts, and matter fields. Because in this
paper we consider only a contribution of the matter superfields,
we present explicit expression only for them:

\begin{equation}\label{PV_Determinants}
\Big(\det PV({\bf V},M)\Big)^{-1} = \int D\Phi\,D\tilde \Phi\,
\exp\Big(i S_{PV}\Big),
\end{equation}

\noindent where (taking into account condition (\ref{K_Fixing}))

\begin{eqnarray}
&& S_{PV}\equiv Z(e,\Lambda/\mu) \frac{1}{4} \int
d^4x\,d^4\theta\, \Big(\Phi^+ e^{\bf V} e^{2V} e^{\bf V} \Phi +
\tilde\Phi^+ e^{-{\bf V}^t} e^{-2V^t} e^{-{\bf V}^t}\tilde\Phi
\Big)
+\qquad\nonumber\\
&& + \frac{1}{2}\int d^4x\,d^2\theta\, M \tilde\Phi^t \Phi +
\frac{1}{2}\int d^4x\,d^2\bar\theta\, M \tilde\Phi^+ \Phi^*.\qquad
\end{eqnarray}

\noindent We assume that $M_i = a_i\Lambda$, where $a_i$ are some
constants. Inserting the Pauli-Villars determinants allows
cancelling the remaining divergences in all one-loop diagrams.

\section{New identity for Green functions and its verification
in the non-Abelian theory} \label{Section_F_Calculation}
\hspace{\parindent}

In the massless case the new identity for Green functions can be
formulated as follows \cite{SD,SDYM}:

It is easy to see that the two-point Green function for the matter
superfield is written as

\begin{equation}\label{Explicit_Green_Functions}
\frac{\delta^2\Gamma}{\delta\phi_x^+\delta\phi_y} = \frac{D_x^2
\bar D_x^2}{16} G(\partial^2) \delta^8_{xy},
\end{equation}

\noindent where $G$ is a function. Then, setting the momentum of
the gauge field to 0, from the Slavnov--Taylor identities it is
possible to find the vertex

\begin{eqnarray}\label{Vertex}
&& \frac{\delta^3\Gamma}{\delta {\bf
V}^a_y\delta\phi^+_{0z}\delta\phi_x}\Bigg|_{p=0} = e \Bigg[-2
\partial^2\Pi_{1/2}{}_y\Big(\bar D_y^2\delta^8_{xy}
\delta^8_{yz}\Big) F(q^2) + \frac{1}{8} D^b C_{bc} \bar
D_y^2\Big(\bar D_y^2\delta^8_{xy} D_y^c \delta^8_{yz} \Big) f(q^2)
-\vphantom{\frac{1}{2}}\nonumber\\ && -\frac{1}{16} q^\mu G'(q^2)
\bar D\gamma^\mu\gamma_5 D_y \Big(\bar D_y^2\delta^8_{xy}
\delta^8_{yz}\Big) -\frac{1}{4} \bar D_y^2\delta^8_{xy}
\delta^8_{yz}\, G(q^2)\Bigg] T^a,
\end{eqnarray}

\noindent where $T^a$ denotes the generators of the gauge group in
a representation, in which the matter superfields are. The
functions $f$ and $F$ can not be found from the Slavnov--Taylor
identities. The new identity can be written in the form

\begin{equation}\label{New_Identity}
\int d^4 q\,\Lambda\frac{d}{d\Lambda} \frac{f(q^2)}{q^2 G(q^2)} =
0.
\end{equation}

\noindent The derivative with respect to $\ln\Lambda$, appearing
in this expression, is introduced in order to obtain well defined
integrals. In the end of this section we explain this by a
concrete example.

In the Abelian case such an identity can be verified by the
straightforward summation of Feynman diagrams \cite{Identity}.
However, the Feynman rules are different in a non-Abelian theory
mostly due to vertexes with the selfaction of the gauge field.
This essentially complicates applying this method. For diagrams,
which do not contain such vertexes the calculations are similar to
the Abelian case. But for diagrams with the triple vertex of the
gauge field a proof, made in Ref. \cite{Identity} is not
applicable or, at least, should be essentially modified. So, there
is a problem, whether the new identity is valid in this case also.
In order to answer it, it is not necessary to calculate all
Feynman diagrams in one or another order of the perturbation
theory. According to Refs. \cite{ThreeLoop,Pimenov}, if we fix an
arbitrary diagram with a loop of the matter superfields and
without external lines, then the new identity should be valid for
the sum of diagrams, which are obtained by cutting a loop of the
matter superfields by all possible ways. (In order to obtain the
vertex function we should attach to them one more line of the
background gauge field by all possible ways.)

\begin{figure}[h]
    \centering
    \includegraphics[scale=0.7]{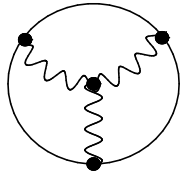}
    \caption{Diagram, generating the considered contribution to the
    new identity}
    \label{ostov}
\end{figure}

In this paper we consider a diagram, presented in Fig.
\ref{ostov}, as a starting point.

\vspace{0.5cm}

\vspace{0.5cm}
\begin{figure}[h]
    \centering
    \includegraphics[scale=0.7]{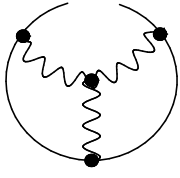}
    \caption{Way of cutting the diagram}
    \label{ostov-cut}
\end{figure}

\begin{figure}[h]
    \centering
    \includegraphics[scale=0.6]{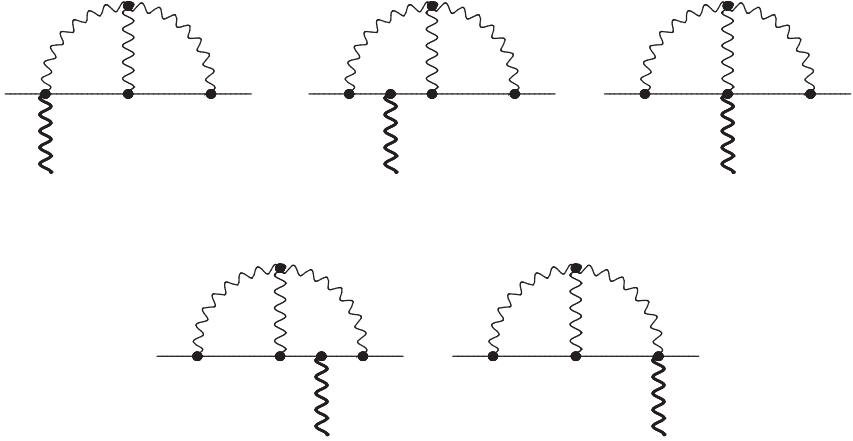}
    \caption{Diagrams, defining the function $f$, corresponding to
    the considered diagrams}
    \label{vertex-diagr}
\end{figure}

\noindent From the topological point of view there is the only way
to cut a loop of the matter superfield, presented in Fig.
\ref{ostov-cut}. Therefore, it is necessary to calculate a set of
diagrams, presented in Fig. \ref{vertex-diagr}. In all these
diagrams the chiral field $\phi$ is at the first external line,
and the non-chiral field $\phi^*_0$ is at the second one.
Therefore, all presented diagrams are not topologically
equivalent.

\vspace{0.5cm}

Calculating these diagrams we can find the function $f$. The
function $G$ in the lowest approximation should be set to 1.
Really, in the tree approximation $G=1$. Therefore, in the given
order for the considered class of diagrams we have:

\begin{equation}
G(q^2) = 1 + O(\alpha^2);\qquad f(q^2) = \alpha^2 f^{(2)}(q^2) +
O(\alpha^3).
\end{equation}

\noindent Therefore,

\begin{eqnarray}\label{Expansion}
\int d^4q\,\Lambda \frac{d}{d\Lambda}\frac{f(q^2)}{q^2 G(q^2)} =
\int d^4 q \,\Lambda \frac{d}{d\Lambda} \frac{\alpha^2
f^{(2)}(q)}{q^2} + O(\alpha^3).
\end{eqnarray}

\noindent So, we see that the considered contribution is actually
determined by the two-loop value of the single function $f^{(2)}$.

In order to find the two-loop value of the function $f^{(2)}$, it
is necessary to make an explicit calculation of Feynman diagrams,
presented in Fig. \ref{vertex-diagr}, using the standard
supergraph technique. The result is (in Euclidean space, after the
Weak rotation)

\begin{eqnarray}\label{f-expression}
&& f^{(2)}(q) = - 2\pi^2\, C_2 \Big( C_2(R) - \frac12\, C_2
\Big)\, \int \frac{d^4k\, d^4l}{(2\pi)^8} \Bigg(
\frac{l^{\mu}}{(k+q+l)^2} + \frac{(k+q)^{\mu}}{(k+q)^2}
\Bigg)\times\\
&& \times \frac{(k+q+l)_{\mu}}{(k+q)^2 \, (k+q+l)^2 k^2 \Big(1 +
k^{2n}/\Lambda^{2n}\Big)\, l^2 \Big(1 + l^{2n}/\Lambda^{2n}\Big)\,
(k+l)^2 \Big(1 + (k+l)^{2n}/\Lambda^{2n}\Big)},\nonumber
\end{eqnarray}

\noindent where $C_2(R)$ and $C_2$ are defined by

\begin{equation}\label{C1-inv}
T^a\, T^a = C_2(R),
\end{equation}
\begin{equation}\label{C2-inv}
f^{amn} \, f^{bmn} = C_2 \, \delta^{ab}.
\end{equation}

\noindent Substituting this expression into Eq. (\ref{Expansion}),
we obtain (technical details are presented in Appendix
\ref{Appendix_Integral}) that in the considered approximation for
the considered diagrams

\begin{eqnarray}\label{Result}
&& \int
\frac{d^4q}{(2\pi)^4}\,\Lambda\frac{d}{d\Lambda}\frac{f(q^2)}{q^2\,G(q^2)}
=\nonumber\\
&& = \alpha^2 \pi^2 \, C_2 \Big( C_2(R) - \frac12\, C_2 \Big)\,
\int \frac{d^4q\, d^4k\, d^4l}{(2\pi)^{12}}
\,\frac{\partial}{\partial q^{\mu}}
\Bigg\{\Lambda\frac{d}{d\Lambda} \, \Bigg[
\frac{(k+q+l)^{\mu}}{q^2 \, (k+q)^2 \,
(k+q+l)^2} \times\qquad \nonumber\\
&& \times \frac{1}{k^2 \Big(1 + k^{2n}/\Lambda^{2n}\Big) \, l^2
\Big(1 + l^{2n}/\Lambda^{2n}\Big)\, (k+l)^2 \Big(1 +
(k+l)^{2n}/\Lambda^{2n}\Big)}\Bigg] \Bigg\} = 0.
\end{eqnarray}

\noindent Therefore, the new identity for Green functions seems to
be valid in the non-Abelian theory.

Using the considered example it is convenient to explain why we
introduce the derivative

\begin{equation}
\Lambda\frac{d}{d\Lambda} = \frac{d}{d\ln\Lambda}
\end{equation}

\noindent in the integrand. Let us first propose that this
derivative is absent. Then, after taking the well defined
integrals with respect to $d^4k$ and $d^4l$ from the dimensional
considerations we obtain the integral

\begin{equation}\label{Integral}
\int \frac{d^4q}{(2\pi)^4} \frac{a(q^2/\Lambda^2)}{q^4},
\end{equation}

\noindent where $a$ is a dimensionless function, which is rapidly
decreasing at $q\to\infty$. In general, it is possible that
$a(0)\ne 0$. (It is easy to see that the value $a(0)$ is a finite
constant.) But if $a(0)\ne 0$, then the integral in Eq.
(\ref{Integral}) is not well defined: it is divergent in the
infrared region. In order to avoid this we introduce the
additional differentiation with respect to $\ln\Lambda$. Due to
its presence the term $a(0)$, which does not depend on $\Lambda$,
disappears, and the integral becomes finite in the infrared
region.

According to Refs. \cite{SD,SDYM} the left hand side of Eq.
(\ref{New_Identity}) is actually a contribution to the two-point
Green function of the gauge field, all other contributions being
integrals of total derivatives. Therefore, appearing of a total
derivative in Eq. (\ref{Result}) confirms a proposal that in
supersymmetric theories all contributions to the Gell-Mann--Low
function are integrals of total derivatives if the higher
derivatives are used for the regularization.


\section{Conclusion}
\label{Section_Conclusion}
\hspace{\parindent}

In this paper we showed that new identity (\ref{New_Identity}) was
also valid in the non-Abelian theory. Similar to the case of the
electrodynamics, it follows from the fact that all integrals
defining the Gell-Mann--Low function are factorized to the total
derivatives. The considered identity seems to be a consequence of
a rather nontrivial symmetry. Deriving this identity from the
first principles is a rather interesting and complicated problem.

Moreover, the calculations, performed in this paper, confirm a
hypothesis that all contributions to the Gell-Mann--Low function
in supersymmetric theories are integrals of total derivatives. The
reason of this fact is also so far unclear.

\bigskip
\bigskip

\noindent {\Large\bf Acknowledgments.}

\bigskip

\noindent This paper was supported by the Russian Foundation for
Basic Research (Grant No. 05-01-00541).

\appendix

\section{Obtaining the integral of total derivative}
\hspace{\parindent}\label{Appendix_Integral}

Here we present a detailed derivation of Eq. (\ref{Result}) from
Eq. (\ref{f-expression}), because it is not quite trivial.

After substituting the function $f$ the left hand side of the new
identity is written as

\begin{eqnarray}\label{Start}
&&  \int d^4 q\,\Lambda\frac{d}{d\Lambda}\frac{f^{(2)}(q)}{q^2} =
- 2\pi^2\, C_2 \Big( C_2(R) - \frac12\, C_2 \Big)
\int \frac{d^4q\, d^4k\, d^4l}{(2\pi)^{12}}
\Lambda\frac{d}{d\Lambda}\times  \nonumber \\
&& \times \Bigg( \frac{l^{\mu}}{(k+q+l)^2} +
\frac{(k+q)^{\mu}}{(k+q)^2} \Bigg) \frac{(k+q+l)_{\mu}}{q^2
(k+q)^2 (k+q+l)^2 k^2 \Big(1 + k^{2n}/\Lambda^{2n}\Big)}
\times\qquad \nonumber\\
&& \times \frac{1}{l^2 \Big(1 + l^{2n}/\Lambda^{2n}\Big)\, (k+l)^2
\Big(1 + (k+l)^{2n}/\Lambda^{2n}\Big)}.
\end{eqnarray}

\noindent In the first term we perform the following sequence of
substitutions: $q \to q-k-l$;\ $k \to -k$;\ $l\to -l$. As a result
we obtain

\begin{equation}
\frac{(k+q+l)_{\mu} l^{\mu}}{q^2 (k+q)^2 (k+q+l)^4} \to -
\frac{q_{\mu} l^{\mu}}{q^4 (q+l)^2 (q+k+l)^2},
\end{equation}

\noindent all other multipliers being the same. Then we perform
the substitutions $l\to l-k$;\ $k \to -k$;\ $k \to l$, after which
this factor becomes

\begin{equation}
- \frac{q_{\mu} (k+l)^{\mu}}{q^4 (q+k)^2 (q+k+l)^2}.
\end{equation}

\noindent And, finally, we add to the expression in the round
brackets in Eq. (\ref{Start})

\begin{equation}
0 = - 2 + \frac{q^{\mu} q_{\mu}}{q^2} + \frac{(k+q+l)^{\mu}
(k+q+l)_{\mu}}{(k+q+l)^2}.
\end{equation}

\noindent Finally, the contribution to the two-point Green
function of the gauge field, we are interested in, can be
rewritten in the form

\begin{eqnarray}
&& \int d^4 q\,\Lambda\frac{d}{d\Lambda} \frac{f^{(2)}(q)}{q^2} =
- 2\pi^2\, C_2 \Big( C_2(R) - \frac12\, C_2 \Big)\, \int
\frac{d^4q\, d^4k\, d^4l}{(2\pi)^{12}}
\Lambda \frac{d}{d\Lambda}\Bigg\{- 2 +\nonumber\\
&& + \frac{q^{\mu}\, (k+q+l)_{\mu}}{q^2} + \frac{(k+q)^{\mu}\,
(k+q+l)_{\mu}}{(k+q)^2} +
\frac{(k+q+l)^{\mu}(k+q+l)_{\mu}}{(k+q+l)^2}
\Bigg\}\frac{1}{(k+q)^2}
\times \qquad\nonumber\\
&& \times \frac{1}{q^2 (k+q+l)^2 k^2 \Big(1 +
k^{2n}/\Lambda^{2n}\Big) l^2 \Big(1 + l^{2n}/\Lambda^{2n}\Big)
(k+l)^2 \Big(1 + (k+l)^{2n}/\Lambda^{2n}\Big)},\qquad
\end{eqnarray}

\noindent Derivation of Eq. (\ref{Result}) from this expression is
evident.


\end{document}